\documentstyle[seceq,supplement,epsf]{ptptex}

\def\bp{{\bf p}}

\let\al=\alpha
\let\be=\beta
\let\ga=\gamma

\let\de=\delta
\let\De=\Delta
\let\ep=\varepsilon

\let\th=\theta

\let\<=\langle
\let\>=\rangle

\def\dsp{\displaystyle}

\let\ad=\dagger
\def\e{ {\rm e} }
\def\beq{\begin{equation}}
\def\eeq{\end{equation}}
\def\ba{\begin{array}}
\def\bea{\begin{eqnarray}}
\def\ea{\end{array}}
\def\eea{\end{eqnarray}}

\def\comment#1{ \hbox{[{\it Comment suppressed here.}\/]} }
\def\hide#1{}

\newsavebox{\eqlabel}

\makeatletter  
\newlength{\numblen}
\newsavebox{\eqnumb}
\def\@eqnnum{\savebox{\eqnumb}{\rm (\theequation)}%
\settowidth{\numblen}{\usebox{\eqnumb}}%
\makebox[\numblen][l]{\usebox{\eqnumb}~~~\usebox{\eqlabel}}}
\makeatother   

\newenvironment{equationwithlabel}[1]{ %
  \savebox{\eqlabel}{#1}
  \begin{equation}\label{#1} }{\end{equation}} 
\newcommand{\beql}[1]{\begin{equationwithlabel}{#1}}
\newcommand{\eeql}{\end{equationwithlabel}}

\notypesetlogo  

\title{
QCD at Finite Baryon Density: Chiral Symmetry Restoration
and Color Superconductivity}

\author{
Krishna {\sc Rajagopal}\footnote{Email address: 
krishna@ctp.mit.edu.  \hfill MIT-CTP-2718}
}

\inst{
Center for Theoretical Physics,\\
Massachusetts Institute of Technology,\\
Cambridge, MA USA 02143
}

\abst{
We use a variational procedure to study finite density QCD
in an approximation in which the interaction
between quarks is modelled by that induced by instantons.
We find that uniform states with
conventional chiral symmetry breaking have negative
pressure
with respect to empty space at all but the lowest
densities, and are therefore unstable.
This is 
a precisely defined phenomenon which motivates 
the basic picture of hadrons assumed in the MIT bag model, with
nucleons as droplets of chiral symmetry restored phase.  
This suggests that the phase transition to a chirally
symmetric phase occurs by percolation of preexisting
droplets of the symmetric phase, and in these proceedings we
expand upon our previous presentation of this observation.  
At all densities high enough that the chirally symmetric phase
fills space, color symmetry is broken
by the formation of a $\langle qq \rangle$ 
condensate of quark Cooper pairs.  A plausible
ordering scheme leads to a substantial gap in a
Lorentz scalar channel involving quarks of 
two colors.
}

\begin{document}

\maketitle

\section{Introduction}

In my lecture at YKIS97, I presented recent work done with
Mark Alford and Frank Wilczek\cite{us}  on QCD at 
finite baryon density.  This contribution to a large
extent follows that
paper, although in Section 3 I have  
given a significantly expanded treatment of chiral symmetry
restoration.

Asymptotic freedom leads us to expect that at high baryon
density quarks
behave nearly freely and form large Fermi surfaces, with 
interactions between the quasiparticles at the 
Fermi surfaces which become weak at asymptotically high
density.
Since the quark-quark interaction is attractive in the color
${\bar {\bf 3}}$ channel, BCS pairing of quarks will
occur no matter how weak the interaction.
Cooper pairs of quarks cannot be color singlets,
and so a $\langle qq \rangle$ condensate inevitably breaks
color symmetry.  This breaking is analogous to the breaking of
electromagnetic gauge invariance in superconductivity.  
In this color superconducting phase, the Higgs mechanism
operates and five of the eight gluons become massive,
and the $U(1)$ gauge boson which remains massless is
a linear combination of the photon and a gluon. 
The proposed symmetry breaking in diquark channels is of course
quite different from chiral symmetry breaking
in QCD at zero density, 
which occurs in color singlet quark-antiquark channels.  

Our goal is to explore such new forms of ordering in the context
of a model
which is definite, qualitatively reasonable, and yet sufficiently
tractable that likely patterns of symmetry breaking and rough
magnitudes of their effects can be identified concretely.  In the
course of looking for new patterns we will need to discuss the fate of
the old one, and here a surprise emerges: we find that a uniform
phase with broken chiral symmetry is unstable at any nonzero density,
and therefore unphysical.
At all but the lowest densities, this instability is signalled by
negative pressure, which presumably triggers the break-up of the
uniform state into regions of high density separated by empty space.
This suggests that in this model, baryons can be identified with 
high baryon density droplets containing massless quarks, and suggests
that chiral symmetry restoration at high density and zero temperature
can be treated as a percolation (or better, as we shall see,
``unpercolation'') process.
In these proceedings, I give an alternate description of 
these phenomena, which has virtues both as a way to
better understand the zero temperature physics of concern
to us here and because it can be extended to nonzero
temperatures.\cite{BR}
We construct the thermodynamic potential for the model at hand,
and discover that it has 
a first order transition between a state
with zero baryon density and  
broken chiral symmetry (namely the vacuum) and a state 
with chiral symmetry restored and with a
baryon density $n_0$ greater than that of nuclear
matter.  
Ordinary
nuclear matter can then be identified within the model
as being in the mixed phase of
this first order transition: it consists of droplets of
the high density phase separated by regions of vacuum.
Chiral symmetry restoration occurs as the first order
transition to the high density phase
is completed as the density is increased and the droplets
merge.
The high density phase is a color superconductor, 
as we discuss in Section 4.\cite{rapp}

\section{Model}


We already alluded to the fact that asymptotic freedom
suggests qualitatively new types of order at very high density.  
This regime has been studied\cite{bailin} by approximating the interquark
interactions by one gluon exchange, which is in fact attractive
in the color antitriplet channel.
Perturbative treatments cannot, by their nature,
do full justice to a problem
whose main physical interest arises at moderate
densities.  To get more insight into the phenomena, and in particular
to make quantitative estimates, it seems appropriate 
to analyze a tractable, physically motivated model.

We have done a variational
treatment of a
two-parameter class of models
having two flavors and three colors of massless quarks.  
The kinetic part of the Hamiltonian is 
that for free quarks, while the
interaction Hamiltonian is a slight idealization
of the  
instanton vertex\cite{thooft} from QCD, explicitly:
\beq
H_I = -\dsp K \int\!d^3x\, 
 \bar\psi_{R1\al}\,{\psi_{Lk}}^\ga \,
 \bar\psi_{R2\be}\,{\psi_{Ll}}^\de \, \ep^{kl} 
 \, (3 {\de^\al}_\ga {\de^\be}_\de - 
 {\de^\al}_\de {\de^\be}_\ga)\ + \ {\rm h.c.}\ ,
\eeq
where $1,2,k,l$ are flavor indices,
$\al,\be,\ga,\de$ are color indices,
repeated indices are summed, and the spinor indices are
not shown. 
The overall sign is chosen negative for later convenience,
so that $K>0$ results in chiral symmetry breaking.
$H_I$ is not yet a good representation of the instanton
interaction in QCD: 
in order to mimic the effects of asymptotic freedom,
we must modify it in such a way that the interaction
decreases with increasing momentum.  We 
write $H_I$ as a mode expansion in momentum
space involving creation and
annihilation operators and spinors, and  
multiply the result by a product of form factors each of the form
$F(p) = [ \Lambda^2 / (p^2 + \Lambda^2) ]^\nu$,
one for each of the momenta of the four fermions.
This
factorized form is taken for later convenience,
and is an idealization.   $\Lambda$, of
course, is some effective QCD cutoff scale, which one might anticipate
should be in the range 300 -- 1000 MeV.  $\nu$ parametrizes the shape
of the form factor; we consider $\nu=1/2$ and $\nu=1$.
Since the  
interaction we have chosen is not necessarily an accurate rendering of 
QCD,
we will have faith only in conclusions that are robust with respect to
the parameter choices.  

The color, flavor, and Lorentz structure of our interaction has been
taken over
directly from the instanton vertex for two-flavor QCD.  
This interaction properly
reflects the chiral symmetry of QCD:
axial baryon number symmetry is broken, while chiral $SU(2)\times
SU(2)$ is respected.  Color 
is realized as a global symmetry.
There are other four-fermion
interactions in addition to $H_I$ which respect the unbroken symmetries
of QCD; 
using $H_I$ alone is the simplest way of breaking all
symmetries broken by QCD, and is therefore a good starting 
point.\footnote{For example, one could add a four-fermion interaction
based on one gluon exchange.\cite{iwado}} 
This model of fermions interacting via a four-fermion interaction
is
one in the long line of such
models inspired by the work of BCS as adapted to particle physics,
starting with Vaks and Larkin\cite{vaks} and 
Nambu and Jona-Lasinio\cite{njl} and studied subsequently by
many others\cite{klevansky}.  There is also a tradition
of modelling the low-energy dynamics of QCD, and specifically the
dynamics of chiral symmetry breaking, with instanton interactions
among quarks derived semi-microscopically.\cite{instantonliquid}  
This approach
cannot explain the other main qualitative aspect of low-energy QCD
dynamics, that is strict confinement of quarks, but it is
adequate for many purposes.

\section{Chiral Symmetry Breaking and Restoration}

We first consider symmetry breaking in the familiar pattern known 
for QCD at zero density.  
Working first at zero density, 
we choose a  variational wave function of
the form
\begin{equation}
\ba{rl}
|\psi\rangle = 
\dsp\prod_{\bp,i,\alpha}
& \Bigl( \cos(\th^L(\bp)) + \e^{i\xi^L(\bp)}\sin(\th^L(\bp))
 a^\ad_{L\,i\alpha}(\bp)  b^\ad_{R\,i\alpha}(-\bp) \Bigr) \\[1ex]
& \Bigl( \cos(\th^R(\bp)) + \e^{i\xi^R(\bp)}\sin(\th^R(\bp))
 a^\ad_{R\,i\alpha}(\bp)  b^\ad_{L\,i\alpha}(-\bp) \Bigr)|0\rangle \ ,
\ea
\label{varwf}
\end{equation}
with variational parameters $\theta$ and $\xi$ depending on the
modes.  $a^\dagger_{i\alpha}$ and $b^\dagger_{i\alpha}$ create particles
and antiparticles respectively, with flavor $i$ and color $\alpha$. 
This standard pairing form preserves the normalization of the
wave function.  The pairing occurs between particles and antiparticles with
the same flavor and color but opposite helicity and  
opposite 3-momentum.
It is in a Lorentz scalar, isospin
singlet component of the chiral $(2,2)$ representation, {\it i.e}. has
the $\sigma$-field quantum numbers standard in this context.
Following a well-trodden path we
find that the energy is minimized when
\beq
\tan(2\th^L(\bp))=\tan(2\th^R(\bp))=\frac{F^2(p) \Delta_\chi}{p}\ ;\ \ \ \ 
\xi^L(\bp) +  \xi^R(\bp) = \pi \ .
\label{variation}
\eeq
The full specification of $\xi^R$ and $\xi^L$ depends on
spinor conventions.
The gap parameter $\Delta$ is momentum independent and
is defined by
\beq
\Delta_\chi \equiv 16 K \int_0^\infty \frac{p^2 dp}{2\pi^2} F(p)
\sin(\theta(p))\cos(\theta(p))\ .
\label{deltadef}
\eeq
(\ref{variation}) and (\ref{deltadef}) are consistent only
if $\Delta=0$ or if $\Delta$ satisfies the gap equation
\beq
1 = 8K \int_0^\infty \frac{p^2 dp}{2 \pi^2} 
{ F^4(p) \over \sqrt{ F^4(p)\Delta_\chi^2 + p^2}},
\label{gapeq}
\eeq
Note that, unlike for standard pairing at a Fermi surface, this
equation does not have solutions for arbitrarily weak coupling, but
only for couplings above a certain threshold value.  
Also note that
although it is common to refer to $\Delta_\chi$ as the gap parameter, 
it is best thought of as inducing an effective quark mass,
which takes the form $\Delta_\chi F^2(p)$.  In the
interests of simplicity of presentation, we quote
all results in this paper for $\Delta_\chi = 400$ MeV;
we have verified that the picture we present is
qualitatively unchanged for $300$ MeV $<\Delta_\chi <$ $500$ MeV.
Fixing $\Delta_\chi$ at zero density
fixes the magnitude (and sign) of the 
coupling $K$ for each $\Lambda$ and $\nu$.  

We now 
generalize this calculation to nonzero quark number 
density $n$.\footnote{The
reader should be warned that the next page or so
is in a sense misleading; as we will see
in due course the state we are about to construct 
is unphysical.}
It is most favorable energetically to fill 
the particle states up
to some Fermi momentum $p_F$, while leaving the
corresponding antiparticle states empty.  
That is, we replace $|0\>$ on the right hand side of (\ref{varwf})
by $|p_F\rangle$, in which all particle states with $|\bp|<p_F$
are occupied.  
The quark number density, 
\beq
n=\frac{2}{\pi^2}\, p_F^3\ ,
\label{ndef}
\eeq
is determined by $p_F$
but as we shall see $p_F$ is not equal to the chemical
potential $\mu$.
The creation operators in (\ref{varwf}) 
for modes below the Fermi surface annihilate $|p_F\>$,
and so for these modes effectively $\theta(\bp)=0$.
On the other hand for
the unoccupied states, with $|\bp| > p_F$, the variational scheme is
unmodified.  Note that the condensate does not
affect $n$, since it pairs quarks and anti-quarks.
Thus we arrive at a very simple modification
of the gap equation: the lower limit of the integral in
(\ref{gapeq})  is $p_F$,
rather than 0.  Since the interaction is ever more effectively 
quenched as $p_F$ increases, the gap
parameter
$\Delta_\chi (p_F)$ arising as the solution of (\ref{gapeq}) 
will shrink monotonically
and eventually vanish as $p_F$ increases.  
Let us define
$n_c$ to be the critical
density at which $\Delta_\chi$ vanishes.

Having obtained a definite wave function, we can evaluate the
energy density in terms of the gap parameter.  Relative to
the energy of the naive vacuum 
state $|0\>$, the energy density $\varepsilon(n)$ is
given by
\beq
{1\over 24}\varepsilon(n) = \frac{p_F^4}{16 \pi^2}+
\int_{p_F}^\infty \frac{p^2 dp}{2 \pi^2}  \frac{p}{2} \left(
 1 - { p \over \sqrt{\De_\chi^2 F^4(p) + p^2}} \right)
 - {\De_\chi^2 \over 32 K}\ ,
\label{chiralen}
\eeq
where $p_F(n)$ is to be obtained from (\ref{ndef}).  
The chemical
potential $\mu$, the minimum energy required to add one more quark
to the state, is given by $\mu=\partial \varepsilon/\partial n$ at fixed 
volume. We have verified that $\mu^2= p_F^2 + \Delta_\chi^2 F^4(p_F)$.
Note that $\varepsilon(0)<0$, reflecting the fact that the
physical vacuum state with its chiral condensate has a lower energy
than the state $|0\>$.  Measured relative to
the physical vacuum, the energy density at nonzero $n$ is
$e(n)\equiv\varepsilon(n)-\varepsilon(0)$.

\begin{figure}[t]
\vspace{-0.55in}
\centerline{
\epsfysize=2.65in
\hfill\epsfbox{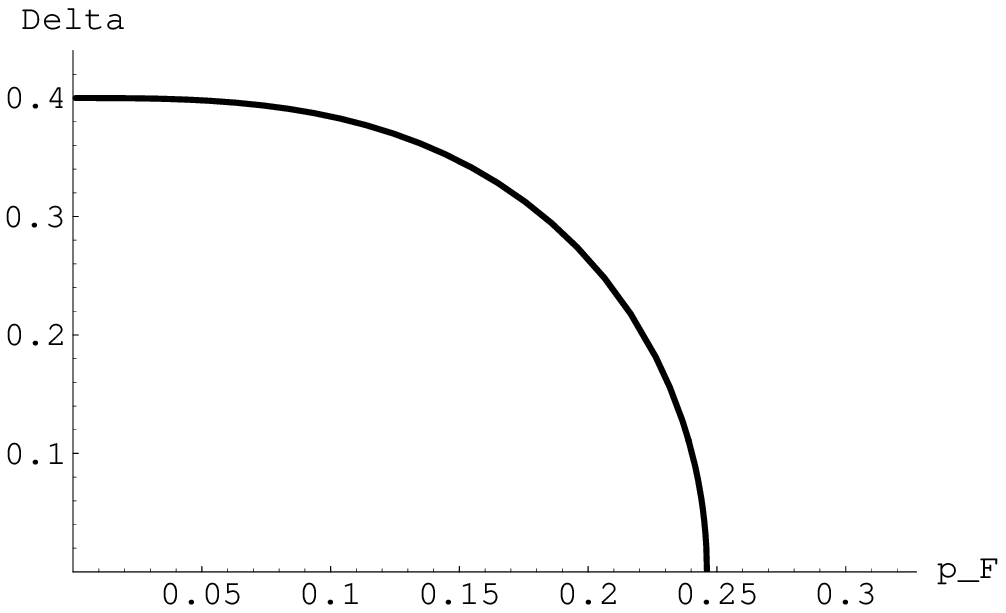}\hfill
\epsfysize=2.65in
\hfill\epsfbox{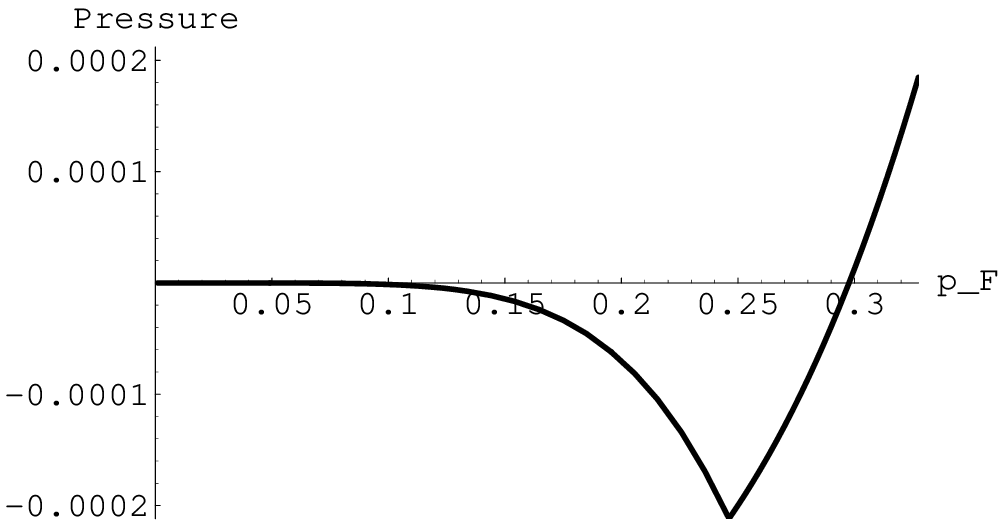}\hfill
}
\vspace{-0.55in}
\caption{Chiral gap (in GeV) and the pressure 
(in GeV$^4$) as a function
of $p_F = (n \pi^2/2)^{1/3}$ in GeV.  
The  
pressure is positive at {\it very}
small $p_F$; it becomes negative at a
$p_F$ which is less than 0.001.
At $n=n_c$,
the gap vanishes and the pressure,
which is still negative,
has a cusp. At $n=n_0>n_c$, the pressure
crosses zero and becomes positive.} 
\vspace{-0.2in}
\end{figure}
The equation for $\Delta_\chi (p_F)$ can be solved numerically, and the
resulting energy evaluated.  Great physical interest attaches to the 
pressure 
\begin{equation}
P(n) ~=~ n \Biggl( {\partial e \over \partial n} 
\Biggr) - e =
n \frac{\partial}{\partial n}\Biggl( \frac{e}{n}
\Biggr)\ 
\end{equation}
of a region with density $n$, where the pressure of the
physical vacuum is by definition zero.
For all values of the parameters that we consider reasonable,
we find that after a tiny interval of very low densities
at which the pressure is positive, {\em the pressure  
becomes negative, and continues to decrease until the
critical density $n_c$}.  At that point we
switch over to an essentially free quark phase, in which
the energy density relative to the physical vacuum
is $3 p_F^4 / (2\pi^2)\, - \,\varepsilon(0)$,
and the pressure is given by $p_F^4/(2\pi^2)\, +\, \varepsilon(0)$. 
At $n=n_c$, where these phases join,
the pressure function is continuous, with a finite
negative value, but it has a cusp.  
As density increases in the
free-quark phase, the pressure increases monotonically, and
passes through zero again at some density $n_0 > n_c$.  
At $n=n_0$, the energy per quark $e/n$ is minimized  and this phase
is stable at zero pressure.
$\Delta$ and $P$ are shown 
for one set of parameters in Figure 1. 


Evidently, at all but the lowest densities (which
we discuss below) in the presence of
a chiral condensate the negative pressure
associated with increasing vacuum energy
overcompensates the increasing Fermi pressure. 
This negative pressure
signals mechanical instability of the uniform
chiral symmetry broken phase.    
There is
an attractive physical interpretation of this phenomenon.  
The
uniform nonzero density phase will break up into 
stable droplets of high
density $n=n_0$ in which the pressure is zero 
and chiral symmetry is restored, surrounded by  
empty space with $n=P=0$.  
Although our simple calculations do not
allow us to
follow the evolution and eventual stabilization of the original quark
cloud, it is hard  
to avoid identifying the droplets of chiral
symmetric phase into which it condenses with physical 
nucleons.  Nothing within the model tells us that
the stable droplets have quark number $3$; nucleons are simply
the only candidates in nature which can be identified with droplets
within which the quark density is nonzero and the chiral condensate
is zero. 
If correct, this identification is very reminiscent of the MIT
bag philosophy, 
here arising in the description of a sharply defined physical
phenomenon.\footnote{Considerations similar
to those we describe also lead Buballa\cite{buballa}
to conclude that 
in a Nambu Jona-Lasinio model
with an interaction which differs from the one we use,
matter with broken chiral symmetry is unstable
and nucleons 
can therefore only be viewed as bags within which chiral symmetry
is restored.}
It seems quite different, at least superficially, from
the Skyrme model, where the chiral symmetry order parameter changes in
direction but not in magnitude within the nucleon.

An alternative way of reformulating the results
just presented is to consider the thermodynamic potential 
$\Omega(\mu,\Delta_\chi)$.
Values of the gap parameter $\Delta_\chi$ which 
solve the gap equation are extrema, satisfying
$\partial \Omega(\mu,\Delta_\chi)/\partial \Delta_\chi=0$,
and at these extrema $\Omega/V = -P = e - \mu n$.
Using a mean field approach to evaluate $\Omega$, the 
following picture emerges.\cite{BR}
At $\mu=0$, $\Omega(0,\Delta_\chi)$ has a
minimum at $\Delta_\chi = 400$ MeV which has $P=\Omega=0$ and  
$n=\partial P/\partial \mu = 0$. This is of course the vacuum
solution to the gap equation.  As $\mu$ is increased sufficiently, 
a second minimum of the potential develops at $\Delta_\chi=0$.
However, the
region around the minimum describing the vacuum state
remains unchanged, and this minimum continues to be the 
lowest possible $\Omega$.
As $\mu$ is increased  further, the value of $\Omega$ at the local minimum
becomes
lower and lower.  The lowest $\Omega$
state of the system remains unchanged (eg the system
is in the vacuum state with density zero and $\Delta_\chi$
unchanged) until the chemical potential 
is increased to $\mu_0 = p_F(n_0)$.  
At $\mu=\mu_0$, 
the local minimum of $\Omega$ at $\Delta_\chi=0$ has $\Omega=0$.
That is, $\Omega(\mu_0,0)=0=\Omega(\mu_0,400{\rm ~MeV})$. 
There is now a $P=\Omega=0$
chirally symmetric phase with density $n_0$.
Before $\mu$ can be increased any further, the system
must undergo a first order phase transition during which
the density increases from zero to $n_0$.  At all intermediate
densities, the system is in a mixed phase in which there are
regions with density $n_0$ and regions with density zero.
Once the transition is completed, $\mu$ can rise further,
and the lowest $\Omega$ configuration becomes one with 
$\Omega<0$, that is with positive pressure, in which
$n>n_0$.  

What, then, of the negative pressure phase constructed
previously?  This solution to the gap equation is a local
{\it maximum} of $\Omega$, with $\Omega>0$ and $P<0$.  Were one
to somehow construct it, it would immediately fall apart into
a mixture of the two stable phases.
The description of the preceding paragraph allows one
to arrive at  the correct description of the 
system (density zero, or density $n_0$, or mixture 
of the two) without ever constructing
an unphysical uniform state 
with $n$ and $\Delta_\chi$ nonzero.

Finally, what 
of the positive
pressure phase at very low density?  
This dilute gas of quarks with mass $\Delta(0)$
has $\mu > \Delta(0) > \mu_0$.  
At this large value of $\mu$, $\Omega$ has 
two nondegenerate minima. The low density gas of constituent
quarks is one; the other 
is a phase
with $n>n_0$, and it is this dense phase
which has the lower $\Omega$.  
The low density gas of constituent quarks 
could only be reached if, as $\mu$ is increased
above $\mu_0$, the phase transition to the favored
high density chirally symmetric phase does not
occur until $\mu$ is increased so high as to be
above $\Delta_\chi(0)$.

The picture we find is satisfying 
since the model finds a way to avoid having
a phase with unconfined constituent quarks even though 
we
have not included confinement 
in any explicit sense.
This nice result is not obtained for all parameter values,
however.
For example, for $\nu=1$ and $\Lambda>2.2$ GeV,  
we find that $\mu_0 > \Delta_\chi$.  That is, 
the vacuum phase becomes a gas of constituent quarks
at a lower $\mu$ than that at which the chirally
symmetric high density phase takes over.
One has a mixed phase of droplets
within which chiral symmetry is restored surrounded
by regions not of vacuum but of regions with 
a nonzero density of constituent quarks.
It is fortunate that these parameter ranges
seem unreasonable.  
 
At a quantitative level, a naive implementation of our proposed
identification of droplets of $n=n_0$ matter with nucleons
works surprisingly well.   One
might want to identify $n_0$ with the quark density at the
center of  baryons.  Taking this to be two(three) times nuclear matter
density yields $p_F(n_0)\sim 0.34(0.39) {\rm ~GeV}$. 
On the other hand, requiring $e(n_0)/n_0$ to be one third
the nucleon mass yields $p_F(n_0)\sim 0.31$ GeV.
Our toy model treatment cannot meet both criteria simultaneously, which
is not surprising,
but the magnitude of $p_F(n_0)$
is very reasonable, ranging from 0.27 to 0.36 for the
parameters we have considered.
The 
vacuum energy, which becomes the bag constant,
is also reasonable in magnitude.
Adding further interactions to $H_I$
would obviously
make a quantitative difference 
but the qualitative picture will not change, at least 
as long as the instanton interaction is the dominant one.

In any case, the physical picture suggested here has significant
implications for the phase transition, as a function of density, to
restored chiral symmetry.   Since the nucleons are regions where the
symmetry is already restored, the transition should occur by
a mechanism analogous to percolation
as nucleons merge.\cite{satz} 
In other words, nuclear matter is midway through the first
order phase transition described above, and chiral symmetry
restoration is nothing more than completing this transition.
This transition should
be complete once a density characteristic of the center of 
nucleons is achieved.  
As $n$ increases, the fraction of the volume of space occupied by the $n=n_0$
phase increases.  In our toy model, the chemical potential and the pressure
remain constant as $n$ is increased
from that of nuclear matter, which is
in the mixed phase, to $n_0$.  This reflects the fact that we have
neglected interactions between droplets. In nature, once
these are included,
some external pressure must
be imposed in order to induce the nucleons to 
merge.\footnote{There is a related reason that 
interactions between droplets are needed.
Although it is very nice to see the
model favoring quarks ``confined'' into droplets within
which they have zero mass rather than free as constituent quarks,
it is
asking too much of this toy model to expect it to correctly
predict that nature favors droplets containing three quarks.
However, augmenting the model to include a
short range repulsion between droplets would
make it unfavorable for
small droplets to merge into bigger ones, and would
mean that in the mixed phase the favored droplets are 
the smallest possible droplets.   A model with  
color introduced only as a global symmetry cannot
select 3-quark-droplets, but it would
nevertheless be interesting to see it favoring small
droplets over large ones.
One way 
of introducing such an effect would be to 
introduce a negative surface tension for the interface
between the $n=n_0$ phase and the vacuum.
With a negative surface tension, in the mixed phase the
system will maximize the ratio of droplet
surface to droplet volume, and will therefore consist 
of the smallest possible droplets of the $n=n_0$ phase
surrounded by vacuum.  
A negative surface tension (or short range droplet-droplet
repulsion introduced in some other way) will not upset the
stability of either the vacuum or of the uniform phase
with density $n=n_0$; it simply upsets the stability
of a mixed 
phase with one large droplet relative to one with many smaller ones.}

The mechanism of chiral symmetry restoration at finite density but
zero temperature is quite different from the one we expect at finite
temperature and zero density: it occurs by percolation 
among pre-formed bags of symmetric phase.  
In the toy model we have described, this transition seems
perfectly smooth.  As the density increases, one phase
steadily wins over the other.  In the chiral limit, however,
this cannot be the whole story.  At zero density, chiral
symmetry is broken. At $n=n_0$ when the dense phase
has taken over completely, chiral symmetry is restored.
At what density (that is at what volume ratio of the two phases in
the mixed phase regime) is chiral symmetry restored?
This question cannot be answered precisely 
using the methods we have developed
to this point. It is plausible, however, that as soon
as the density is high enough that there are no longer
vacuum regions of infinite spatial extent, chiral
symmetry is restored. Once one has only isolated vacuum
regions, each of these isolated regions can have the
chiral order parameter pointing in different directions
in isospin space.   Each isolated region will pick a 
different $\sigma$ direction.  This means that in the
system as a whole, chiral symmetry will be restored.
In this sense, what matters is the ``unpercolation'' of
the vacuum rather than the percolation of the droplets.
Regardless of the precise
criterion, the density $n_\chi$ at which chiral symmetry restoration
occurs must be somewhat less than $n_0$. 
At densities less than $n_\chi$, one has
massless pions --- oscillations of the orientation of the condensate
present in the vacuum regions of infinite spatial
extent associated with spontaneous chiral symmetry breaking.  
At higher densities $n>n_\chi$, the pions need no longer be massless.
It is a logical possibility that the restoration of 
chiral symmetry at $n=n_\chi$ is a second order transition,
occuring within the ``larger'' first order transition.
It is also a logical possibility that the 
chiral symmetry restoration transition is first order,
between two phases with density $n_\chi^-$ and $n_\chi^+$,
again within the larger first order transition between
$n=0$ and $n=n_0$.  Of course, away from the chiral limit,
with a nonzero current quark mass and a nonzero pion
mass, the transition can be completely smooth.

\section{Color Superconductivity}

At high density, pairing of particles near the Fermi surface as in the
original BCS scheme becomes more favorable.  
Our Hamiltonian supports
condensation in quark-quark channels.   
The condensation is now between fermions with the same
helicity, and the Hamiltonian selects chiral isosinglets --- that is,
antisymmetry in flavor.  
One can therefore have spin 0 --- antisymmetric in spin and therefore
in color, forming a $\bar {\bf 3}$, or spin 1 --- symmetric in spin and
therefore in color, forming a {\bf 6}.   

We first consider the former.  A suitable trial wave 
function is\footnote{Nothing misleading here; there will
be no negative pressure and no
droplet formation here; the state we are about to
construct is stable.}
\beq
|\psi\> = G_L^\ad G_R^\ad |p_F\>
\eeq
where
\beq
\ba{rcll}
G_L^\ad &=& \dsp
 \prod_{\alpha,\beta,\bp}
& \Bigl( \cos(\th^L_{A}(\bp)) + \ep^{\alpha\beta 3}\e^{i\xi^L_A(\bp)}
\sin(\th^L_{A}(\bp))
 a^\ad_{L\,1\alpha}(\bp)  a^\ad_{L\,2\beta}(-\bp) \Bigr) \\[1ex]
&&& \Bigl( \cos(\th^R_{B}(\bp)) + \ep^{\alpha\beta 3}\e^{i\xi^R_B(\bp)}
\sin(\th^R_B(\bp))
 b^\ad_{R\,1\alpha}(\bp)  b^\ad_{R\,2\beta}(-\bp) \Bigr) \\[1ex]
&&& \Bigl( \cos(\th^R_{C}(\bp)) + \ep^{\alpha\beta 3}\e^{i\xi^R_C(\bp)}
\sin(\th^R_{C}(\bp))
 a_{R\,1\alpha}(\bp)  a^\ad_{R\,2\beta}(-\bp) \Bigr) \\[1ex]
G_R^\ad &=& \multicolumn{2}{l}{\hbox{same, with~~}R\leftrightarrow L}.
\ea
\label{colwf}
\eeq
Here,
$\alpha$ and $\beta$ are color indices, and we
have chosen to pair quarks of the first two colors.
$1$ and $2$ are flavor indices. The first term in (\ref{colwf})
creates particles above the Fermi surface; the second creates
antiparticles; the third creates holes below the Fermi surface.
In this state, the Lorentz scalar
$\langle q^{i\alpha} \,C\gamma^5 q^{j\beta}\varepsilon_{ij}\,
\varepsilon_{\alpha\beta 3}\rangle$ is nonzero. This 
singles out a preferred
direction in color space and breaks color $SU(3) \rightarrow SU(2)$.
The $U(1)$ of electromagnetism is spontaneously broken but
there is a     linear combination of electric charge
and color hypercharge under which the condensate is neutral, and
which therefore generates an unbroken
$U(1)$ gauge symmetry.  No flavor symmetries,
not even the chiral ones,  are broken.

Note that $n$ is now not given by (\ref{ndef}) because
the operators in (\ref{colwf}) can change particle number.
Varying the expectation value of $H-\mu N$ in this state
with respect to $p_F$ yields $p_F=\mu$, unlike in 
the case of the chiral condensate.  This difference
reflects the fact that 
a gap in a $\<qq\>$ channel does not act as an effective mass
term in the way that a chiral gap                  
does.  Upon adding a quark, the condensate can adjust in such
a way that the energy cost is only $p_F$.   There
is, however, a true gap in the sense of
condensed matter physics: the energy cost of 
making a particle-hole excitation is  
$2\Delta F^2(\mu)$ at minimum.
Varying with respect to 
all the other variational parameters yields
$\xi^R_{A,B,C} + \xi^L_{A,B,C} = \pi$, $\theta^R_{A,B,C}=
\theta^L_{A,B,C}$, and
\beq
\tan(2\theta^L_A(\bp)) = \frac{F^2(p)\Delta}{p-\mu}\ ,
\ \ \
\tan(2\theta^L_B(\bp)) = \frac{F^2(p)\Delta}{p+\mu}\ ,
\ \ \
\tan(2\theta^L_C(\bp)) = \frac{F^2(p)\Delta}{\mu-p}\ .
\label{colorvariation}
\eeq
Here, the gap $\Delta$ satisfies
a self-consistency equation of the form
\begin{eqnarray}
1 = \frac{K}{\pi^2} \biggl\{ 
\int_\mu^\infty { F^4(p) p^2 dp 
\over\sqrt{ F^4(p)\De^2 + (p-\mu)^2}}
&+& \int_0^\infty 
{ F^4(p) p^2 dp \over \sqrt{ F^4(p)\De^2 + (p+\mu)^2}}\nonumber\\
&+&\int_0^\mu 
{F^4(p) p^2 dp \over \sqrt{ F^4(p)\De^2 + (\mu-p)^2}}\biggr\}\ .
\label{colorgapeq}
\end{eqnarray}
The three terms in this equation arise respectively from 
particles above the 
Fermi surface, antiparticles, 
and particles below the Fermi surface.  
For $\mu >  0 $ the particle and hole integrals diverge logarithmically
at the Fermi surface 
as $\Delta \rightarrow 0$, which signals the possibility of
condensation for arbitrarily weak attraction.


Because the numerical  coefficient in
the gap equation is smaller than
the threshold value at which one would
have a nonzero $\Delta_\chi$ at $\mu=0$, $\Delta$ would be zero were
it not for the logarithmically divergent contribution
to the integral from the region near $\mu$. 
This means that at small $\mu$, the gap must be small
because the density of
states at the Fermi surface is small.  This has only formal
significance, because the only densities of physical relevance
are $n=0$ and $n\ge n_0$. At intermediate densities, matter
is in an inhomogeneous mixture of the $n=0$ and $n=n_0$
phases. 
As $\mu$ increases, the density of states at
the Fermi surface increases and the gap parameter grows.  Finally, at
large $\mu$ the effect of the form factor $F$ is felt,
the effective coupling decreases, and the
gap parameter goes back down.  For the parameter ranges we have examined
the gap parameter is quite substantial:  $\sim 50-150$ 
MeV at $n_0$, and
peaking at $100-200$ MeV at a density somewhat higher. We plot
$\Delta$ for two sets of parameters in Figure 2.  The density
at which the gap peaks depends on $\Lambda$; the shape
of the curve depends on $\nu$; the height of the curve is
almost independent of both.
\begin{figure}[t]
\vspace{-0.55in}
\centerline{
\epsfysize=2.65in
\hfill\epsfbox{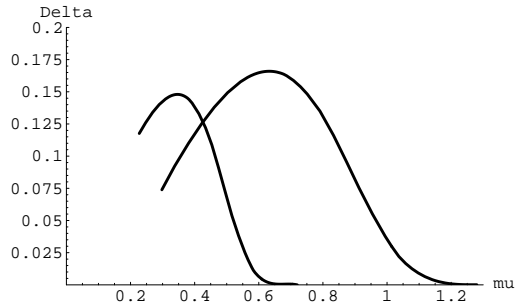}\hfill
}
\vspace{-0.55in}
\caption{Gap created
by the Lorentz scalar color superconductor 
condensate,
as a function of $\mu=p_F$ for $\nu=1$ and (from left to right) 
$\Lambda=0.4,0.8$ GeV. Each curve begins
where $n$ is given by the appropriate $n_0$.}
\vspace{-0.2in}
\end{figure}

As in the previous section, 
one can obtain expressions for the energy and the density,
and thus derive the equation of state.  We find that the equation of
state is hardly modified from the free-quark values --- the
presence of the condensate typically lowers the
energy density by several tens of MeV per fm$^3$ and the pressures,
at equal baryon density, are 
equal to within a few per cent.  This makes it very plausible that 
the color condensation makes only a small change in the 
density $n_0$ at which a stable phase exists at zero pressure. 
We have made this argument more rigorous by doing a 
a calculation in which we consider chiral and
color condensation simultaneously.\cite{BR}  

The third color has so far been left out in the cold, but we can gain
energy by allowing it too to condense.  The available channel is the
color {\bf 6}.  There is attraction in this channel,\cite{us} for a
pairing that is a spatial axial vector.  
The resulting condensate
spontaneously breaks both color and rotational symmetry.
Without modification, our model suggests that this axial
vector gap is of order a few keV at most,\cite{us} orders of
magnitude smaller then the scalar color breaking condensate.
This is not a robust result, and it is worth
exploring whether plausible interactions can be added to $H_I$ which
strengthen the axial vector condensate.

In looking for signatures of color superconductivity in heavy ion physics
and in neutron stars, it is unfortunate that the equation
of state is almost equal to that for a deconfined phase
with no diquark condensate.
Superconducting condensates do modify the gauge interactions and
this may have implications in heavy ion collisions.
In our model as it stands, color 
is realized as a global symmetry.  Breaking of this symmetry generates
Nambu-Goldstone bosons, formally.  
However,
in reality color
is of course
a gauge symmetry, and the true spectrum does not contain massless
scalars, but rather massive vectors.   
There is 
a residual $SU(2)$ gauge symmetry,
presumably deconfined,
and there are five gluons whose mass is set by the scalar 
condensate.   
The scalar condensate carries electric as well as color charge.  It
is neutral under a certain combination of electrodynamic and color
hypercharge.   
If densities above $n_0$ are achieved at
low enough temperatures that the scalar condensate forms,
there will
be a mixture of the ordinary photon and the color hypercharge gauge
boson which is massless.
Either the modification of the photon
or the loss of massless gluons could 
have consequences, but dramatic effects
do not yet seem apparent.  
Recently, we have described another
possible signature.\cite{us2} The formation of a color
superconducting state lowers the energy most if the matter
is up-down symmetric. In a heavy ion collision, in which
down quarks initially exceed up quarks, as a color superconductor
condensate forms in the densest region near the center of
the collision it may therefore expel a few
down quarks and up anti-quarks into the surrounding
less dense regions. These will eventually become
negative pions. When the condensate breaks up late in
the collision, it likely yields equal numbers of protons
and neutrons.  The total number of protons in the final
state will then be more than twice 79.
Experimentalists should therefore look for
event-by-event 
correlation between observables which suggest high density and
low temperature on the one hand, and either an increase in
the number of protons per event or an increase in the $\pi^-/\pi^+$
ratio on the other.

Turning to neutron 
star phenomenology, the slowness
of observed neutron star cooling rates may indicate 
that a gap in the excitation
spectrum for quark matter might be welcome,\cite{ogleman}  
as this suppresses neutrino emission via weak interaction
processes involving single
thermally excited $u$ and $d$ quarks by $\exp(-\Delta/T)$.

\section{Discussion}

Many things were ignored in this analysis.  Most important, perhaps,
is the strange quark.  
Whatever the interaction(s), color superconductivity in
a three flavor theory necessarily introduces the new
feature of flavor symmetry breaking. The 
condensates considered in our two-flavor model are flavor singlets;
this is impossible for a $\langle qq \rangle$ condensate
in a three flavor theory.  One particularly attractive
possibility
is condensation in the 
$\langle q^\alpha_i \, C \gamma^5 q^\beta_j \varepsilon^{ijA}
\varepsilon_{\alpha\beta A}\rangle$
channel, with summation over $A$.
This breaks 
flavor and color in a coordinated fashion,
leaving unbroken the
diagonal subgroup of $SU(3)_{\rm color}\times SU(3)_{\rm flavor}$.

Another question concerns the postulated Hamiltonian.  While there are
good reasons to take an effective interaction of the instanton type
as a starting point, there could well be 
significant corrections affecting the more delicate
consequences such as axial vector {\bf 6} condensation.   A specific,
important example is to compare the effective interaction derived from
one-gluon exchange.   It turns out that this interaction has a
similar pattern, for our purposes, to the instanton: it is very attractive in
the $\sigma$ channel, attractive in the color antitriplet scalar, and
neither attractive nor repulsive in the color sextet axial vector.
More generally, it would be desirable to use a renormalization
group treatment to find the interactions which are
most relevant near the Fermi surface.

The qualitative model we have treated suggests a compelling picture
both for the chiral restoration transition and for the color
superconductivity which sets in 
at densities just beyond. It
points toward future
work in many directions:  chiral
symmetry restoration via the percolation (of baryons)
or unpercolation (of the vacuum) 
must be further characterized; consequences in neutron 
star and heavy ion physics remain to be elucidated;
the superconducting ordering patterns
may hold further surprises, particularly
as flavor becomes important.

\section*{Acknowledgements}

I would like to thank Tsuneo Suzuki and the other organizers
of YKIS 97, first for arranging such a stimulating
workshop, and second for giving me the opportunity
to visit Japan for the first time, which I very
much enjoyed.
I am very grateful to my collaborators M. Alford, 
F. Wilczek and J. Berges.  Exploring this terrain
with them has been great fun. 
I also wish to acknowledge helpful conversations
with S. Pratt, who suggested the possibility of
negative surface tension, and with 
M. Gyulassy, R. L. Jaffe, M. Prakash, R. Rapp, 
H. Satz, T. Sch\"afer, 
R. Seto, E. Shuryak, D. T. Son, M. Stephanov, M. Velkovsky
and I. Zahed. 
This research is supported in part by DOE cooperative
research agreement DE-FC02-94ER40818.

\end{document}